# Secondary Phase Limited Metal-Insulator Phase Transition in Chromium Nitride Thin Films

Bidesh Biswas[1,2]‡, Sourjyadeep Chakraborty[1,2] ‡, Anjana Joseph[1,2], Shashidhara Acharya[1,2], Ashalatha Indiradevi Kamalasanan Pillai[3], Chandrabhas Narayana[1,4], Vijay Bhatia[3], Magnus Garbrecht[3] and Bivas Saha[1,2,5]

[1]*Chemistry and Physics of Materials Unit, Jawaharlal Nehru Centre for Advanced Scientific Research, Bangalore 560064, India.*
[2]*International Centre for Materials Science, Jawaharlal Nehru Centre for Advanced Scientific Research, Bangalore 560064, India.*
[3]*Australian Centre for Microscopy and Microanalysis, The University of Sydney, Camperdown, NSW 2006, Australia.*
[4]*Rajiv Gandhi Centre for Biotechnology, Poojappura, Thiruvananthapuram, 695014, India.*
[5]*School of Advanced Materials, Jawaharlal Nehru Centre for Advanced Scientific Research, Bangalore 560064, India.*

Chromium nitride (CrN) is a well-known hard coating material that has found applications in abrasion and wear-resistant cutting tools, bearings, and tribology applications due to its high hardness, high-temperature stability, and corrosion-resistant properties. In recent years, CrN has also attracted significant interest due to its high thermoelectric power factor, and for its unique and intriguing metal-insulator phase transition. While CrN bulk single-crystals exhibit the characteristic metal-insulator transition accompanied with structural (orthorhombic-to-rocksalt) and magnetic (antiferromagnetic-to-paramagnetic) transition at ~260-280 K, observation of such phase transition in thin-film CrN has been scarce and highly debated. In this work, the formation of the secondary metallic $Cr_2N$ phase during the growth is demonstrated to inhibit the observation of metal-insulator phase transition in CrN thin films. When the Cr-flux during deposition is reduced below a critical limit, epitaxial and stoichiometric CrN thin film is obtained that reproducibly exhibits the phase transition. Annealing of the mixed-phase film inside reducing $NH_3$ environment converts the $Cr_2N$ into CrN, and a discontinuity in the electrical resistivity at ~ 277 K appears which supports the underlying hypothesis. A clear demonstration of the origin behind the controversy of the metal-insulator transition in CrN thin films marks significant progress and would enable its nanoscale device realization.

Correspondence to be addressed to bsaha@jncasr.ac.in and bivas.mat@gmail.com
‡ These authors contributed equally to this work.

# 1. INTRODUCTION

Transition metal mono-nitrides (TMNs) are an emerging class of materials and have attracted significant interest in recent years for their applications in thermoelectric energy conversion, plasmonics, refractory nano-electronic, and optoelectronic devices [1-6]. CrN is one of the most celebrated TMN due to its corrosion-resistant high hardness that is utilized in developing medical implants, silver-lustre in decorative coating, abrasion/wear-resistant drill bits, cutting tools, and other tribology applications [7-11]. It was further demonstrated that CrN has a high thermoelectric power factor (~ $1.7 \times 10^{-3}$ W/m-K$^2$) at 400°C that could be useful for waste-heat recovery and thermoelectric cooling applications [12-15]. However, unlike almost all other TMNs, CrN exhibits an interesting and unusual first-order electronic, magnetic, and structural phase transition at ~ 260 - 280 K that has been researched extensively for both fundamental physics studies as well as technological applications [16-21].

At low temperatures (≤ Neel temperature ($T_N$) ~ 260 - 280 K), CrN is antiferromagnetic in nature with an orthorhombic (Pnma) structure, but above $T_N$, it transforms into a paramagnet with a rocksalt (Fm$\bar{3}$m) structure. Concomitant with the magnetic and structural transition, CrN also undergoes a metallic-to-insulating electronic phase transition when the temperature is increased above $T_N$ ~ 260-280 K [17,22-24]. The origin behind this transition is theoretically predicted to be the antiferromagnetic ordering of Cr atoms below the Neel temperature that induces anisotropic stress and closes the Fermi gap, thus making the CrN metallic [25,26]. This anisotropic stress breaks the symmetry of the rocksalt structure, transforming the crystal structure into an orthorhombic symmetry. Such antiferromagnetic ordering in CrN could be useful for secure data storage and memory devices.

Though the first-order phase transition in bulk CrN has been observed on more than one occasion from the early 1960s, CrN thin films do not always exhibit such phase transition for reasons presumed to be due to (a) variations in stoichiometry, (b) presence of strain in CrN thin films referred as epitaxial constraints, and (c) occurrence of defects such as nitrogen-vacancy, oxygen impurities, etc. [27-31] Early molecular beam epitaxy (MBE) deposited CrN thin film on (001) MgO substrates exhibited a metal-semiconductor electronic transition at ~ 260 - 285 K with a hysteresis of about 20 K [32]. Subsequent in-situ scanning tunneling microscopy (STM), and variable low-temperature reflection high energy electron diffraction (RHEED) studies also revealed a structural transition, while temperature-dependent neutron scattering revealed a magnetic transition [24,29]. The same magnetic transition, however, was absent in CrN films deposited with RF-plasma assisted MBE on (001) MgO and (0001) Al$_2$O$_3$ substrates. Rather interestingly, the CrN/Al$_2$O$_3$ exhibited ferromagnetic-like behaviour resembling the magnetic nature found in Cr-doped dilute magnetic semiconductors. Sputter-deposited CrN$_{1-x}$ thin films with $x \leq 0.03$ deposited on (001) MgO substrate exhibited semiconducting behaviour for a wide temperature range without any discontinuity in the resistivity prompting the authors to conclude that CrN could be a Mott-Hubbard insulator with correlation energy driving open an energy gap of 0.7 eV measured with optical absorption studies [27,33]. Subsequent reports on CrN thin films in some cases exhibited a discontinuity in the resistivity representative of a phase transition in polycrystalline and/or epitaxial films, while on other occasions no transition was observed [34-36]. Similar to the controversy of the presence/absence of the phase transition, the magnitude of the resistivity of the CrN films also vary by more than six orders of magnitude with the lowest resistivity ranging from a few mΩ-cm to about kΩ-cm at room temperature [16,32,37]. Theoretically, first-principles modelling has predicted that CrN could be close to a borderline charge-transfer insulator with a gap between the *3d*-states of Cr and *2p*-states of N [38,39].

In summary, the metal-insulator electronic, orthorhombic-rocksalt structural, and antiferromagnetic-paramagnetic magnetic phase transition in CrN is a subject of intense debate and controversy for more than two decades. Since most studies involving the physics of the phase transition and its practical use in devices require reproducible CrN phase transition in thin films, it is necessary that a suitable growth methodology is developed to observe the phase transition, and to determine the factors that control and impact it. In this letter, we show that the presence of secondary metallic $Cr_2N$ phase, which usually gets incorporated in CrN thin films at a higher Cr-flux rate during depositions (magnetron sputtering in the present case), inhibits the observation of electronic phase transition in CrN thin films. When the Cr-flux is reduced during the deposition below a critical limit, phase-pure single-crystalline and stoichiometric CrN thin films are achieved that reproducibly exhibit the first-order phase transition. In addition, we find that the change in the concentration of $Cr_2N$ inside CrN alters the resistivity by nearly three orders of magnitude which could explain the variation of resistivity in the literature. Our results also show that annealing the $Cr_2N$/CrN mixed lateral heterostructure in a reducing environment such as $NH_3$ at high temperature (800°C) transforms the $Cr_2N$ into CrN and results in a clear discontinuity in the electrical resistivity at ~ 277 K, representative of its metal-insulator phase transition.

## 2. EXPERIMENTAL METHODS

*Sample Growth:* CrN films were deposited on 10×10 mm2 single-crystal MgO (001) substrates using a DC reactive magnetron sputtering system with a base pressure of ~ 2 × 10-9 Torr. The growth temperature and pressure were set to 800°C and 10 mTorr respectively. High-pure Ar (99.99999%) and N2 (99.99999%) gas with a constant Ar:N2 ratio of 9:6 was used for all the depositions. The film stoichiometry was controlled by changing the Chromium (99.95% pure) target power to 100 W, 50W, and 25 W for higher (HCF), medium (MCF), and low (LCF) metal Cr flux films respectively. Prior to deposition, the MgO substrate was cleaned with Acetone and Methanol inside an ultrasonic bath and then thermally treated at growth temperature for 1 hour. For a fixed growth duration of 4 hours, film thicknesses for the HCF, MCF, and LCF were 980, 480, and 240 nm respectively. Such large film thickness minimizes the possibility of epitaxial strain and its impact on the phase transition phenomena.

*X-Ray Diffraction and Raman Analysis:* The high-resolution X-ray diffractogram measurements were carried out with the Bruker D8 system. The Cu anode equipped with Ni filter was used for Cu-Kβ free X-ray source. Raman Spectroscopic measurements were performed with a confocal micro-Raman spectrometer (LABRAM HR Evolution) in the 1800 backscattering mode, and at grazing incidence with the 532 nm excitation from Nd:YAG solid state laser, and an 1800 gr/mm grating. The temperature-dependent Raman studies were carried out using a liquid nitrogen cooled temperature stage (Linkam THMS 600) equipped with a temperature controller (Linkam TMS 94) having a temperature accuracy of ± 1 K. The maximum laser power density was 1.5 mW mm-2 at the sample surface with 400 s of acquisition time, which was low enough to avoid any noticeable local heating of the sample.

*SEM Analysis*: SEM imaging was carried out using FEI Inspect F50. Plan- view SEM images of the films were taken at 15 kV with a secondary electron detector.

*TEM and TKD Analysis:* HRTEM and TKD analysis were carried out at the Australian Centre for Microscopy and Microanalysis, University of Sydney. For sample preparation, a 100 nm Pt protective cap was deposited at the target location with a 5 kV electron beam followed by a 1 µm Pt+C protective cap with a 12 kV Xe beam in Thermo Fisher Scientific Helios Hydra PFIB. Trenching and lift out were done with a 30 kV Xe beam at 60, 15, 4, and 1 nA. Then the sample was welded to Mo grid using a 30 kV Xe beam Pt weld. Thinning was done with tilt angles of ±1.5° with currents of 300, 100, and 30 pA, checking for electron transparency using a 5 kV electron beam with a secondary electron detector. When the ROI was thin enough it was polished with a 5 kV Xe beam with tilt angles of ±3.5°. Final cleaning was performed with a 2 kV Xe beam with tilt angles of ±5.5°.

The HRSTEM images and EDS maps were recorded with an image- and probe-corrected and monochromated Themis-Z 60-300 kV equipped with a high-brightness XFEG source and Super-X EDS detector system for ultra-high-count rates, operated at 300kV. For STEM imaging and EDX mapping, the probe corrector was used to form a focused probe of 0.7 Å diameter. The Super-X EDX detector system enables the recording of high spatial resolution EDS maps.

TKD was performed on a Thermofisher Helios Hydra PFIB using an Oxford Symmetry EBSD detector. Maps were collected at 30 kV with approximately 3.2 nA beam current. A collection speed of 12 ms per point was used and a step size of 10 nm. A pre-tilted holder was used with a tilt angle of -20°, the stage was forward-tilted to 20° to provide an incident beam perpendicular to the sample.

*Electrical Resistivity Measurement:* The electrical resistivity was measured in standard four-probe geometry using cryogen free quantum device ppms. versalab system.

*Thermal Annealing:* Thermal annealing of the HCF-CrN was performed in an NH3 environment inside a tube furnace at 800°C for 12 hrs. Before heating, the whole system was purged with $NH_3$ for 30 minutes to partially remove residual gas impurities. The gas flow was controlled with a flow controller and was kept constant.

## 3. RESULTS AND DISCUSSIONS

Symmetric $2\theta$-$\omega$ HRXRD diffractogram of the films shows (see Fig. S1 in SI) the presence of predominant rocksalt 002 orientation of the CrN on (001) MgO substrates on all three films. All three 002 diffraction peaks are located at ~ 43.6° corresponding to a rocksalt CrN lattice constant of 4.15 Å, which is close to its's bulk value [10,32]. However, along with the rocksalt diffraction peak, the HCF film exhibits an additional peak at ~ 40.3°, which is not present in the other two films. Careful analysis reveals the peak at 40.3° stems from the hexagonal $Cr_2N$ phase with 0002 orientation having a *c*-axis lattice constant of 4.48 Å, which is consistent with literature reports [40,41]. Therefore, it is clear from the HRXRD measurements that while the LCF film exhibits only rocksalt diffraction peak, indicative of its single rocksalt phase nature, the HCF film exhibits both the rocksalt CrN and hexagonal $Cr_2N$ phases. Though the MCF film does not show the presence of the $Cr_2N$ peak, as we shall see later, it contains a small concentration of $Cr_2N$ grains that are not crystallographically aligned for the 0002 Bragg diffraction in HRXRD. It is important to note that the full-width-at-the-half-maxima (FWHM) of the rocking curve ($\omega$-scan) corresponding to the 002 rocksalt CrN peak in the LCF film exhibits a small value of 0.38°, which is representative of its excellent crystalline quality and epitaxial single-crystalline nature.

To obtain the thickness, size, and orientations of $Cr_2N$ and CrN grains, band contrast images (see Fig. 1(a)) and Euler colour maps were obtained from the TKD data (see Fig. 1(a-d)). As seen in Fig. 1(a) $Cr_2N$ appears darker in contrast compared to the CrN, and establishes a dense interconnected metallic network. Though the initial growth of the film on (001) MgO substrate in most places starts with the CrN phase, at some regions V-shaped $Cr_2N$ phase grows on CrN pyramidal side faces. As the film grows, $Cr_2N$ grains appear almost at random points but always with a distinct boundary with the CrN. Pole figure images for the rocksalt {100} set of planes show that the rocksalt phase grows with the [001] (001) CrN ∥ [001] (001) MgO epitaxial relationship on MgO substrate predominantly, marked by the clustering of pixels at the centre and at the corners (with an ~ 6 degree offset in Fig. 1(c) appearing due to tilt in the sample preparation). The corresponding rocksalt regions are shown with maroon and purple colours in the Euler maps with both representing (001) CrN with respect to the c-axis and are interchangeable in a cubic structure, merely indicating a 90° difference in the indexing of the a and b-axes. Similarly, Fig. 1(d) representing pole figures for the {0001} orientations of the hexagonal $Cr_2N$ phase show that most of the hexagonal grains are not aligned parallel to the *c*-axis of the rocksalt CrN, rather makes multiple orientations. Four of such orientations are found predominantly, which are oriented approximately 55° away from the z-axis in the four different quadrants. TKD analysis suggests that the HCF and MCF film contain about 59.8 and 16.7 vol. % $Cr_2N$ respectively.

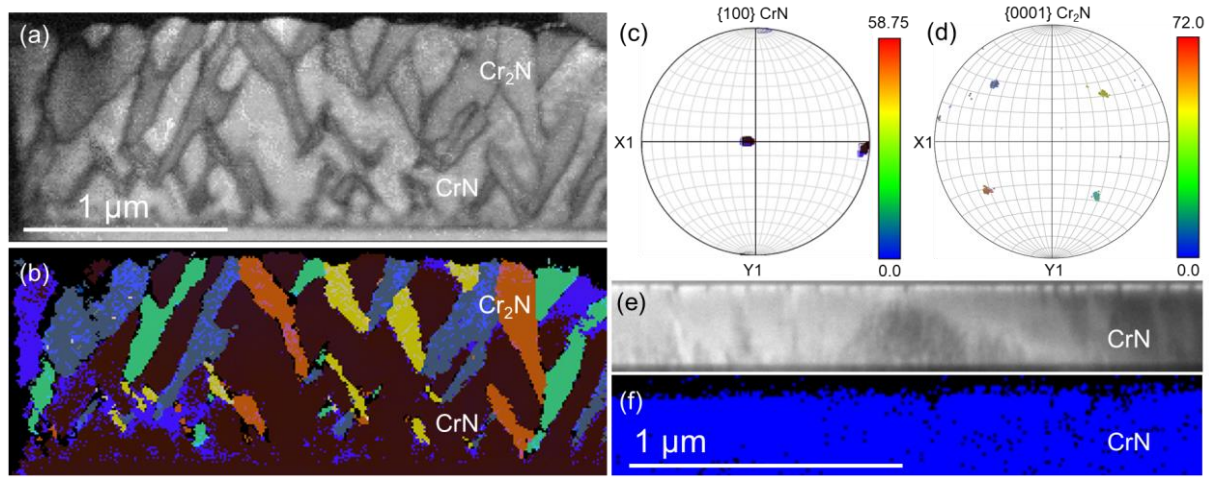

Figure 1(a) Transmission Kikuchi Diffraction (TKD) band contrast mapping of the HCF film is presented. Dark contrasted regions are hexagonal $Cr_2N$ phase, while the light-contrast regions are rocksalt CrN. (b) Euler colour mapping of the HCF film grains along with pole figures in (c) and (d) which show the orientation of different grains inside the film. (e) Band contrast TKD image of the LCF film showing the presence of single-phase CrN grains is shown. (f) TKD phase mapping of the LCF film verifying the single-phase rocksalt CrN growth.

In contrast to the higher and medium flux film, band contrast (Fig. 1(e)) and Euler map (Fig. 1 (f)) of the LCF film shows the presence of only the rocksalt phase and (001) orientation aligned with the substrate, which is consistent with the HRXRD analysis.

While the TKD analysis is used to obtain the phase and crystal orientations with spatial resolutions of ~ 10-15 nm, aberration-corrected HR(S)TEM imaging is employed to determine the microstructure at the atomic level. Low-magnification high-angle annual dark-field scanning transmission electron microscopy (HAADF-STEM) imaging (see Fig. 2(a)) of the HCF film shows the presence of both $Cr_2N$ and CrN inside the film with the former appearing brighter. The V-shaped grains close to the substrate show a strong contrast due to the atomic number (Z) sensitivity of HAADF-STEM. While both the grains start to grow at random locations inside the film, it is interesting to note that they reach the surface approximately at the same height, thus making the surface appear smooth with a feature size of ~ 500-600 nm (also seen in plan-view SEM image in Fig. S3 in SI). A few voids (dark contrast) at the intersection between different grains are also observable in the image. The electron diffraction pattern (EDP) collected from the substrate and the film region in Fig. 2(a)), and presented in the inset, shows both the ordered cubic and hexagonal symmetries in different orientations. STEM-energy dispersive X-ray spectroscopy (EDS) mapping analysis (see Fig. 2(b)) clearly shows the $Cr_2N$ regions with a higher concentration of Cr atoms. Quantification of the grains confirms the CrN and $Cr_2N$ stoichiometry within 1 atomic % accuracy. A HAADF-STEM image (Fig. 1(c)) of the mixed-phase regions as well as individual and combined Cr, N EDS maps are further presented (Fig. 1(c-f)) to show the spatial distribution of the different grains.

HAADF-STEM image of the LCF film appears homogeneous and uniform with only the CrN phase present, as shown in Fig. 2(g). The EDP presented at the inset also verifies the presence of rocksalt CrN only inside the film. STEM-EDS elemental mapping also confirms the homogeneous distribution of Cr and N atoms throughout the film. The CrN/MgO interface image is further presented that shows cubic epitaxial CrN growth on MgO substrate. Due to about ~ 1.4 % lattice mismatch between the CrN and MgO, misfit dislocations are visible at the interface.

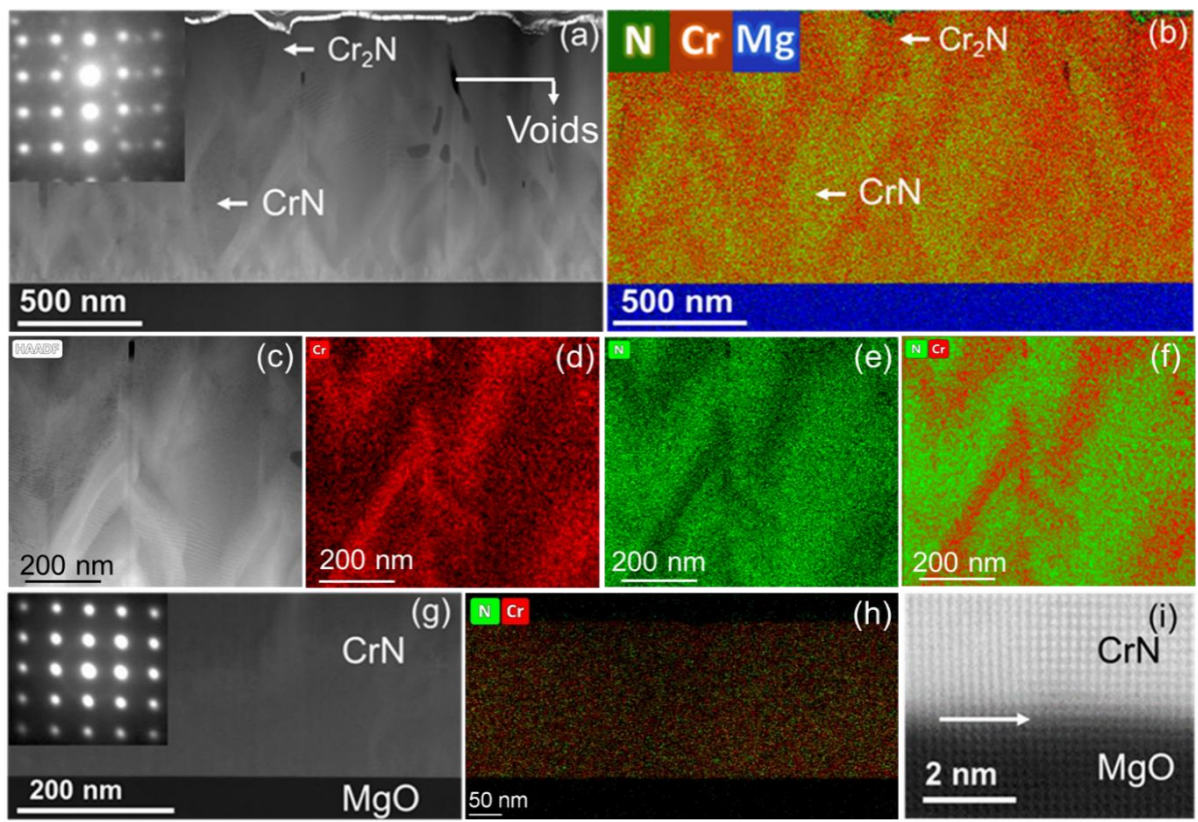

Figure 2 (a) Low-magnification HAADF-STEM micrograph of the HCF film is presented that shows the presence of both $Cr_2N$ and CrN. Inset shows the EDP of the film with cubic [001] and hexagonal patterns. (b) STEM-EDS elemental maps corresponding to (a) show the spatial distribution and morphology of the $Cr_2N$ and CrN grains. (c) HAADF-STEM micrograph of the HCF film with bright regions that represent the $Cr_2N$, while the darker parts
6

showing CrN grains. (d) Cr, (e) N and (f) Cr+N STEM-EDS maps corresponding to the region (c) showing the different grains. (g) HAADF-STEM micrograph of the LCF film showing single-phase epitaxial single-crystalline CrN growth on MgO substrate. (h) Homogeneous and uniform Cr and N atomic distribution in the LCF film is demonstrated by the STEM-EDS map. (i) Atomic-resolution STEM image of the CrN/MgO interface from the LCF film is presented that exhibits cubic epitaxial CrN crystal growth on MgO substrate.

Room-temperature Hall mobility and temperature-dependent electrical resistivity of the films are measured with a 4-point probe method inside a vacuum chamber from 50 K to 400 K. At room temperature, the HCF, MCF, and LCF film exhibit electrical resistivity of 0.6, 3.6, and 200 mΩ-cm respectively, highlighting an increased resistivity for the LCF film (single-phase CrN) by nearly three orders of magnitude with respect to the film having the highest amount of $Cr_2N$, i.e. HCF film. Similarly, the carrier concentration and mobility of the three films are measured to be $4\times10^{21}$ cm$^{-3}$ and 3 cm$^2$/V-s, $7\times10^{20}$ cm$^{-3}$ and 0.98 cm$^2$/V-s, and $1.8\times10^{19}$ cm$^{-3}$ and 3.5 cm$^2$/V-s respectively. The metallic-like high carrier concentration of the HCF and MCF films are presumed to be due to the presence of metallic $Cr_2N$ inside the CrN matrix that increases the carrier concentration and decreases mobility. The carrier concentration of $1.8\times10^{19}$ cm$^{-3}$ of the LCF film is representative of its semiconducting nature with carriers appearing from native defects such as nitrogen vacancies and oxygen impurities during the deposition process.

Both the HCF and MCF films exhibit (see Fig. 3) a metallic-like increase in resistivity with the increase in temperature, and with no apparent discontinuity in the resistivity across the entire 50 – 400 K temperature range. The first derivative of the resistivity with respect to the temperature ($\frac{d\rho}{dT}$), however, showed an abrupt change at ~ 277 K, which is close to the expected first-order phase transition of CrN film. This change is due to the fact that both the HCF and MCF films contain 40.2% and 83.3% rocksalt CrN respectively, which presumably undergoes a metal-insulator phase transition at ~ 277 K, but their effects get suppressed by the change in electrical resistivity of the metallic $Cr_2N$. The change in $\frac{d\rho}{dT}$ is more pronounced in the MCF film as it contains a lower content of $Cr_2N$ compared to the HCF film. The resistivity of the LCF film clearly shows a discontinuity at ~ 277 K representative of its electronic phase transition. Above the transition temperature, the resistivity decreases with an increase in temperature that represents the semiconducting nature of CrN, while at low temperature, the resistivity exhibits a rather interesting U-shaped behaviour with the decrease in temperature that arises from electronic weak localization effects. Detailed analysis on the mechanism of electronic transport at low and high temperatures, however, is beyond the scope of the present study and will be presented subsequently. The derivative of the resistivity vs temperature indicates that the electronic transition indeed occurs in the HCF and MCF films; however, the large content of metallic $Cr_2N$, dominates the transport behaviour of the films. Suitable reduction of Cr-flux below the critical limit leads to the stoichiometric rocksalt CrN that exhibits the metal-insulator transition.

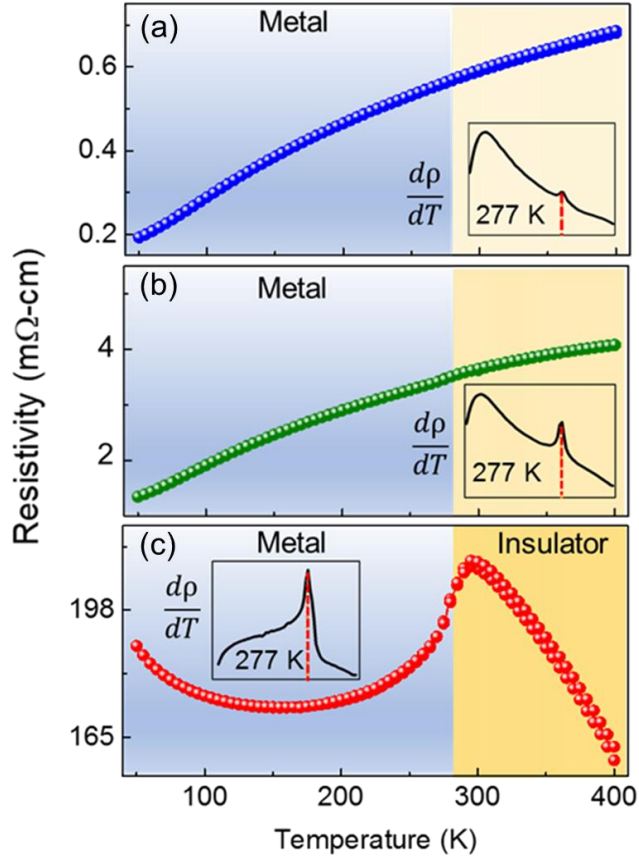

Figure 3 (a) Temperature-dependent electrical resistivity of the HCF film is presented, which shows a metallic-like increase in resistivity with an increase in temperature. The inset showing the $\frac{d\rho}{dT}$ exhibits a small discontinuity at ~ 277 K. (b) The MCF film also exhibits metallic-like increase in resistivity with a more pronounced $\frac{d\rho}{dT}$ discontinuity at the expected phase transition temperature. (c) The LCF film exhibits a well-pronounced clear metal-insulator electronic phase transition at ~ 277 K, with the inset showing the large change in $\frac{d\rho}{dT}$.

While the temperature-dependent electrical measurements demonstrate the electronic phase transition in single-phase stoichiometric CrN single-crystalline film, Raman spectroscopy is used to observe the changes in phonon modes across the temperature induced structural phase transition. Orthorhombic (Pnma) CrN unit cell contains 8 atoms and as a result, there are 24 phonon modes that can be categorized as $4A_g$, $2B_{1g}$, $4B_{2g}$, and $2B_{3g}$ Raman active modes, $2A_u$ silent modes, and $4B_{1u}$, $2B_{2u}$, and $4B_{3u}$ infrared active modes. On the other hand, the two atoms of the rock salt (Fm$\bar{3}$m) CrN unit should exhibit three acoustic modes with zero frequency and three optical modes having $T_{1u}$ symmetry. As the crystal structure changes from rock salt-to-Orthorhombic when the temperature is lowered, the $T_{1u}$ mode is expected to split into $B_{1u}$, $B_{2u}$, and $B_{3u}$ modes. [45] Since the infrared-active and Raman-active modes are exclusive to each other in centrosymmetric structures, only the Raman active Gerade modes can be seen in Raman measurements.

Temperature-dependent Raman measurements have been carried out on the HCF and LCF films and the representative Raman spectra at various temperatures are shown in Fig. 4(a) and (b), respectively. At 300K, CrN exhibits rock salt crystal structure, and as a result, the first-order Raman scattering is symmetry forbidden. Raman spectrum at 300K, therefore, does not exhibit any peak for the LCF film. Also, the Raman spectrum at ambient temperature does not seem to change with the incorporation of secondary phase $Cr_2N$. Even though the rock salt structure does not give rise to any Raman scattering, usually the presence of point defects relaxes the inversion symmetry and defect-induced peaks are visible in several rock salt materials, especially on transition metal nitrides such as ScN, TiN, [42-44] etc. Therefore, the complete absence of any Raman activity in rock salt CrN at 300K highlights its excellent crystalline quality with very little defect densities that could otherwise have relaxed the symmetry rules. With a decrease in temperature, CrN undergoes rock salt ($Fm\bar{3}m$) to orthorhombic ($P_{nma}$) structural phase transition and as a result, various Raman modes start to appear. The onset of transition is observed from 265 K for the HCF film and from 288 K for the LCF film with the appearance of orthorhombic $A_g(1)$ mode. This vibrational mode could be observed at 223 cm$^{-1}$ that starts to appear around the phase transition temperature and becomes intense at lower temperatures. The shoulder peak at 219 cm$^{-1}$ corresponds to $B_{1g}$ mode also shows similar temperature-dependent behaviour as that of $A_g(1)$ mode. Another small peak at 291 cm$^{-1}$ corresponds to the second $A_g$ mode. The peaks at 469 cm$^{-1}$ and 517 cm$^{-1}$ have very small Raman intensity and these could correspond to $A_g(3)$ and $B_{1g}(2)/A_g(4)$ respectively. All the experimentally observed phonon modes have been assigned based on the factor group analysis and Raman measurements by Quintela *et al.* [13] However the authors could not distinguish between $B_{1g}$ and $A_g$ modes as they could not observe the splitting of these modes. In our study, the strong Raman signal coming from the pure phase CrN shows the sharp and intense nature of $B_{1g}$ and $A_g$ modes and their clear splitting, which also demonstrates the highly crystalline quality of the film. The structural transition temperature of the LCF film, as seen in the Raman measurement, is preceding the electronic transition temperature observed in electrical resistivity study as Raman Spectroscopy is very sensitive to subtle structural changes. The transition is less sharp, and the peaks are weaker and broader in the case of the HCF film which could be due to the smaller grain size arising from the high Cr-flux. The splitting of modes is also not seen in the HCF film due to the reduced Raman signal resulting from the grain size effect. New Raman modes will only appear after a sufficient number of the grains have changed to an orthorhombic structure and hence the structural transition occurs at a relatively low temperature in the HCF film. Temperature-dependent Raman spectroscopy analysis, therefore, gives evidence for the structural phase transition occurring both in LCF and HCF films. Our study thus clearly establishes that the structural transition accompanies the electronic transition in stoichiometric pure phase CrN.

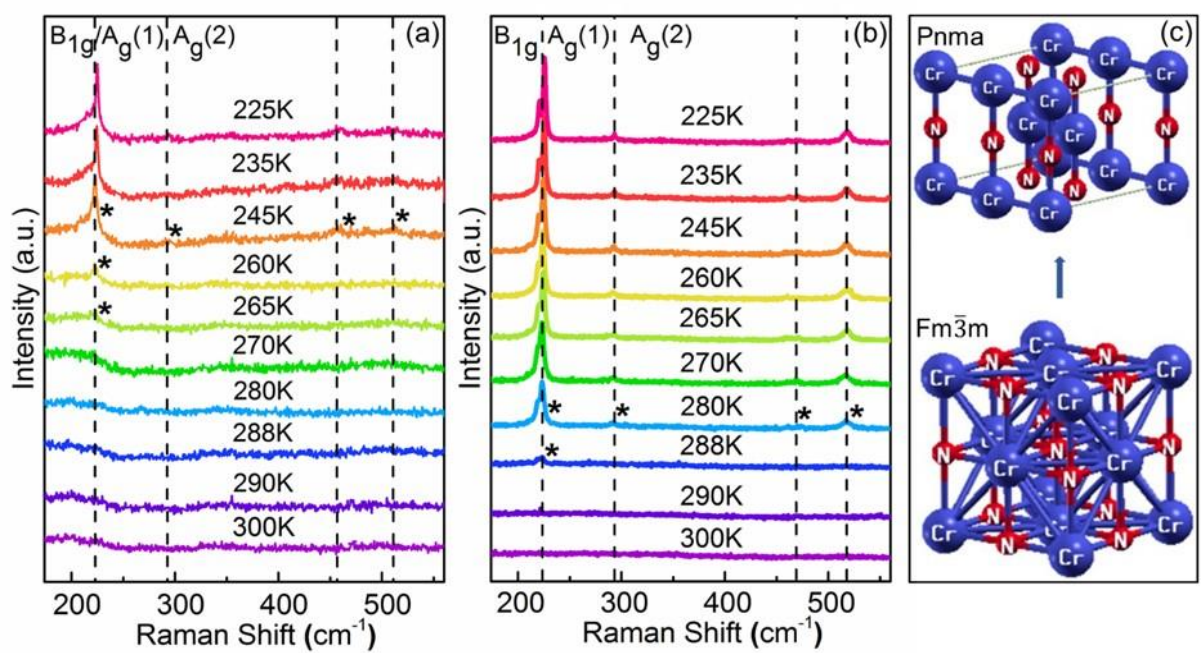

Figure 4. (a) Temperature-dependent evolution of Raman spectra of the HCF thin film and (b) Temperature-dependent evolution of Raman spectra of the LCF thin film showing that at 300K, rock salt CrN do not exhibit any Raman peaks. As the temperature is reduced below the Neel temperature, both the films undergo rock salt-to-orthorhombic structural phase transition and Raman modes start to appear (new modes are shown with black coloured asterisks). The phase transition temperature is seen to be suppressed in the HCF film. (c) Orthorhombic and rocksalt unit cells of CrN are presented to show their structural symmetry.

To demonstrate further that the metal-insulator electronic phase transition can be recovered by removing $Cr_2N$, the HCF film is annealed inside reducing $NH_3$ atmosphere at 800°C for 12 hours (see methods for experimental details). $NH_3$ transforms $Cr_2N$ into CrN as shown in the following chemical reaction.

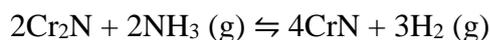

$$2Cr_2N + 2NH_3 (g) \leftrightarrows 4CrN + 3H_2 (g)$$

Structural characterization of the annealed HCF film with HRXRD shows (See Fig. S2 in SI) only the 002 diffraction peak, representative of its rocksalt crystal structure. The hexagonal 0002 $Cr_2N$ peak that appeared in the non-annealed film completely disappeared. The HRTEM imaging of the annealed HCF film shows (see Fig. 5 (a)) the presence of single phase semiconducting CrN only. Unlike the unannealed film, the EDP pattern only shows the diffraction spot corresponding to rocksalt CrN only. Besides, the STEM-EDS mapping (Fig. 5 (b) & (c)) exhibits the uniform distribution of elements throughout the film. However, as a result of prolonged annealing at elevated temperature, several void/pores which are also visible.

Temperature-dependent electrical resistivity measurement shows that unlike the HCF film, the annealed HCF film shows a clear discontinuity ~ 277 K that is representative of its electrical phase transition. However, unlike the LCF film, above the Neel temperatures, the resistivity increases with an increase in temperature that represents its degenerate semiconducting or semi-metallic nature. The creation of defects in high concentration, mainly the nitrogen vacancies, and/or the left out metallic $Cr_2N$ grains in the annealing process possibly results in such degenerate semiconducting or semi-metallic nature in the annealed film. Hall measurements confirm a much high carrier concentration of $4.6 \times 10^{20}$ cm$^{-3}$ in the annealed HCF film with respect to the carrier density of $1.9 \times 10^{19}$ cm$^{-3}$ in the LCF film. Nevertheless, the removal of $Cr_2N$ in the film recovers the metal-insulator phase transition that was inhibited in the HCF film. The present analysis conclusively demonstrates that the presence of secondary metallic $Cr_2N$ phase in CrN inhibits the observation of its metal-insulator phase transition.

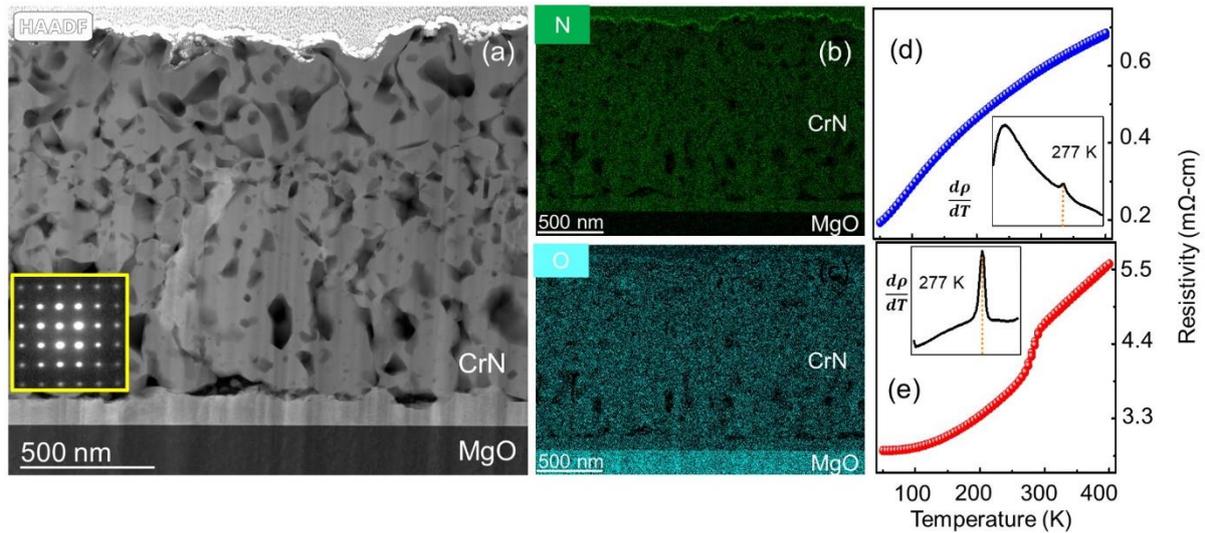

Figure 5. (a) HAADF-STEM micrograph of the annealed film is presented, which shows the presence of single phase semiconducting CrN only. Several voids/pours are visible in the film as a result of prolonged annealing. STEM-EDS elemental mapping of N (b), and O (c) in the annealed film. N is uniformly distributed in the sample. Temperature-dependent resistivity the film before (e) and after (f) the annealing process. The discontinuity in resistivity ~277K represent the electronic phase transition in the annealed film.

## 4. CONCLUSION

In summary, we show that incorporation of the secondary metallic $Cr_2N$ inside CrN during the thin film growth process inhibits the observation of its metal-insulator electronic phase transition. Detailed spectroscopic and microscopic characterization with high-resolution X-ray diffraction, transmission Kikuchi diffraction, and high-resolution transmission electron microscopy reveals that a high Cr-flux during the deposition process results in metallic $Cr_2N$ networks inside the CrN matrix that suppress its electronic phase transition. When the Cr-flux is reduced below the critical limit of the growth rate of 1 nm/min, a stoichiometric single-crystalline CrN thin film is obtained that reproducibly exhibits the electronic phase transition. The structural rocksalt-to-orthorhombic phase transition is also observed with temperature-dependent Raman measurements for phase pure CrN thin films as well as in the presence of metallic $Cr_2N$ in the film. Our results also show that when the $Cr_2N$, inside the high-flux mixed film, is converted to CrN by annealing it inside $NH_3$ environment at high temperature, the electronic phase transition is recovered. This strongly manifests the fact that phase pure and stoichiometric CrN films can show the metal-insulating electronic phase transition. Thus, our results mark an exemplary progress in thin-film CrN research and answer the origin behind the controversy of its electronic phase transition.

## ACKNOWLEDGEMENT

BB, SC, and BS acknowledge support from the International Centre for Materials Science and Sheikh Saqr Laboratory of the Jawaharlal Nehru Centre for Advanced Scientific Research. BS acknowledges the Young Scientist Research Award (YSRA) from the Board of Research in Nuclear Sciences (BRNS), Department of Atomic Energy (DAE), India with grant number 59/20/10/2020-BRNS/59020 for financial support. The authors acknowledge Sydney Microscopy and Microanalysis at the The University of Sydney.

**Conflict of Interest**

The authors declare no conflict of interest.

**Supporting Information**

Supporting Information contains information related to the HRXRD, SEM and TEM-EDX characterization of the films.

**References**

[1] U. Guler, V. M. Shalaev and A. Boltasseva, Nanoparticle plasmonics: going practical with transition metal nitrides, Mater. Today, 18 (2015) 227-237.
[2] G. V. Naik, J. Kim and A Boltasseva, Oxides and nitrides as alternative plasmonic materials in the optical range, Opt. Mater. Express, 1 (2011)1090-1099.
[3] P. Eklund, S. Kerdsongpanya, and B. Alling, Transition-metal-nitride-based thin films as novel energy harvesting materials, J. Mater. Chem. C, 4 (2016) 3905-3914.
[4] M. Khazaei, M. Arai, T. Sasaki, C. Y. Chung, N. S. Venkataramanan, M. Estili, Y. Sakka, and Y. Kawazoe, Novel Electronic and Magnetic Properties of Two-Dimensional Transition Metal Carbides and Nitrides, Adv. Funct. Mater. 23 (2013) 2185-2192.
[5] B. Anasori, M. R. Lukatskaya, and Y. Gogotsi, 2D metal carbides and nitrides (MXenes) for energy storage, Nat. Rev. Mater. 2 (2017) 16098.
[6] H. O. Pierson, "Handbook of Refractory Carbides and Nitrides: Properties, Characteristics, Processing and Applications", Noyes Publications, New York, 1996.
[7] G. Berg, C. Friedrich, E. Broszeit, and C. Berger, Development of chromium nitride coatings substituting titanium nitride, Surf. Coat. Technol. 86-87 (1996) 184-191.
[8] J. A. Sue, and T. P. Chang, Friction and wear behavior of titanium nitride, zirconium nitride and chromium nitride coatings at elevated temperatures, Surf and Coat. Technol. 76-77 (1995) 61-69.
[9] M. Sikkens, A. A. M. T. V. Heereveld, E. Vogelzang, and C. A. Boose, The development of high performance, low cost solar-selective absorbers, Thin Solid Films 108 (1983) 229-238.
[10] P. Hones, N. Martin, M. Regula, and F. Lévy, Structural and mechanical properties of chromium nitride, molybdenum nitride, and tungsten nitride thin films, J. Phys. D: Appl. Phys. 36 (2003) 1023.
[11] M. Lackner, W. Waldhauser, L. Major, and M. Kot, Tribology and Micromechanics of Chromium Nitride Based Multilayer Coatings on Soft and Hard Substrates, Coatings 4 (2014) 121-138.
[12] CX Quintela, F. Rivadulla, and J. Rivas, Thermoelectric properties of stoichiometric and hole-doped CrN, Appl. Phys. Lett. 94 (2009) 152103.
[13] C. X. Quintela, J. P. Podkaminer, M. N. Luckyanova, T. R. Paudel, E. L. Thies, D. A. Hillsberry, D. A. Tenne, E. Y. Tsymbal, G. Chen, C. B. Eom, F. Rivadulla, Epitaxial CrN Thin Films with High Thermoelectric Figure of Merit, Adv. Mater. 27 (2015) 3032-3037.
[14] A. L. Febvrier, N. V. Nong, G. Abadias, and P. Eklund, P-type Al-doped Cr-deficient CrN thin films for thermoelectrics, Appl. Phys. Express 11 (2018) 051003.
[15] M. N. Gharavi, S. Kerdsongpanya, S. Schmidt, F. Eriksson, N. V. Nong, J. Lu, B. Balke, D. Fournier, L. Belliard, A. Le. Febvrier, C. Pallier, P. Eklund, Microstructure and thermoelectric properties of CrN and CrN/Cr2N thin films, J. Phys. D: Appl. Phys. 51 (2018) 355302.

[16] J. D. Browne, P. R. Liddell, R. Street, and T. Mills, An investigation of the antiferromagnetic transition of CrN, Phys. Status Solidi A 1 (1970) 715-723.

[17] L. M. Corliss, N. Elliott, and J. M. Hastings, Antiferromagnetic Structure of CrN, Phys. Rev. 117 (1960) 929.

[18] R. M. Ibbersion, R. Cywinski, The magnetic and structural transitions in CrN and (CrMo)N, Physica B: Condensed Mattter 180-181 (1992) 329-332.

[19] F. Rivadulla, M. B. Lopez, CX Quintela, A. Pineiro, V. Pardo, D. Baldomir, M. A. L. Quintela, J. Rivas, C. A. Ramos, H. Salva, J. S. Zhou, J. B. Goodenough, Reduction of the bulk modulus at high pressure in CrN, Nat. Mater. **8** (2009) 947-951.

[20] I. Stockem, A. Bergman, A. Glensk, T. Hickel, F. Kormann, B. Grabowski, J. Neugebauer, B. Alling, Anomalous Phonon Lifetime Shortening in Paramagnetic CrN Caused by Spin-Lattice Coupling: A Combined Spin and *Ab Initio* Molecular Dynamics Study, Phys. Rev. Lett. 121 (2018) 125902.

[21] S. Zhang, Y. Li, T. Zhao, Q. Wang, Robust ferromagnetism in monolayer chromium nitride, Sci. Rep. 4 (2014) 5241.

[22] A. Mrozinska, J. Przystawa, and J. Sólyom, First-order antiferromagnetic transition in CrN, Phys. Rev. B 19 (1979) 331.

[23] P. A. Bhobe, A. Chainani, M. Taguchi, T. Takeuchi, Ritsuko Eguchi, M. Matsunami, K. Ishizaka, Y. Takata, M. Oura, Y. Senba, H. Ohashi, Y. Nishino, M. Yabashi, K. Tamasaku, T. Ishikawa, K. Takenaka, H. Takagi, S. Shin, Evidence for a Correlated Insulator to Antiferromagnetic Metal Transition in CrN, Phys. Rev. Lett. 104 (2010) 236404.

[24] K. Alam, S. M. Disseler, W. D. Ratcliff, J. A. Borchers, R. P. Pérez, G. H. Cocoletzi, N. Takeuchi, A. Foley, A. Richard, D. Ingram, A. R. Smith, Structural and magnetic phase transitions in chromium nitride thin films grown by rf nitrogen plasma molecular beam epitaxy, Phys. Rev. B 96 (2017) 104433.

[25] A. Filippetti, and N. A. Hill, Magnetic Stress as a Driving Force of Structural Distortions: The Case of CrN, Phys. Rev. Lett. 85 (2000) 5166.

[26] A. Filippetti, W. E. Pickett, and B. M. Klein, Competition between magnetic and structural transitions in CrN, Phys. Rev. B 59 (1999) 7043.

[27] D. Gall, C-S. Shin, R. T. Haasch, I. Petrov, and J. E. Greene, Band gap in epitaxial NaCl-structure CrN (001) layers, J. Appl. Phys. 91 (2002) 5882.

[28] X. Y. Zhang, J. S. Chawla, R. P. Deng, and D. Gall, Epitaxial suppression of the metal-insulator transition in CrN, Phys. Rev. B 84 (2011) 073101.

[29] A. Ney, R. Rajaram, S. S. P. Parkin, T. Kammermeier, and S. Dhar, Magnetic properties of epitaxial CrN films, Appl. Phys. Lett. 89 (2006) 112504.

[30] X. F. Duan, W. B. Mi, Z. B. Guo, and H. L. Bai, A comparative study of transport properties in polycrystalline and epitaxial chromium nitride films, J. Appl. Phys. 113 (2013) 023701.

[31] Y. Tsuchiya, K. Kosuge, S. Yamaguchi, and N. Nakayama, Mater. Trans., Non-stoichiometry and Antiferromagnetic Phase Transition of NaCl-type CrN Thin Films Prepared by Reactive Sputtering, JIM 37 (1996) 121-129.

[32] C. Constantin, M. B. Haider, D. Ingram, and A. R. Smith, Metal/semiconductor phase transition in chromium nitride (001) grown by rf-plasma-assisted molecular-beam epitaxy, Appl. Phys Lett. 85 (2004) 6371.

[33] X. Y. Zhang, and D. Gall, CrN electronic structure and vibrational modes: An optical analysis, Phys. Rev. B 82 (2010) 045116.

[34] Q. Jin, H. Cheng, Z. Wang, Q. Zhang, S. Lin, M. A. Roldan, J. Zhao, J. Wang, S. Chen, M. He, C. Ge, C. Wang, H. B. Lu, H. Guo, L. Gu, X. Tong, T. Zhu, S. Wang, H. Yang, K.J. Jin, E. J. Guo, Strain-Mediated High Conductivity in Ultrathin Antiferromagnetic Metallic Nitrides, Adv. Mater. 33 (2021) 2005920.

[35] Q. Jin, Z. Wang, Q. Zhang, J. Zhao, H. Cheng, S. Lin, S. Chen, S. Chen, H. Guo, M. He, C. Ge, C. Wang, J. O. Wang, L. Gu, S. Wang, H. Yang, K. J. Jin, E. J. Guo, Structural twinning-induced insulating phase in CrN (111) films, Phys. Rev. Materials 5 (2021) 023604.

[36] Zhang, X. Y., J. S. Chawla, B. M. Howe, and D. Gall, Variable-range hopping conduction in epitaxial CrN (001), Phys. Rev. B 83 (2011) 165205.

[37] Z. Hui, X. Zuo, L. Ye, Z. Wang, Z. Zhu, Solution Processable CrN Thin Films: Thickness-Dependent Electrical Transport Properties, Materials, 13 (2020) 417.

[38] P. S. Herle, M. S. Hegde, N. Y. Vasathacharya, S. Philip, M. V. R. Rao, and T. Sripathi, Synthesis of TiN, VN, and CrN from Ammonolysis of $TiS_2$, $VS_2$, and $Cr_2S_3$, J. Solid State chem. 134 (1997) 120-127.

[39] A. Herwadkar, and WRL Lambrecht, Electronic structure of CrN: A borderline Mott insulator, Phys. Rev. B 79 (2009) 035125.

[40] T. Cheiwchanchamnangij, and WRL Lambrecht, Quasiparticle self-consistent G W band structure of CrN, Phys. Rev. B 101 (2020) 085103.

[41] G. Gubert, R. C. Oliveira, D. S. Costa, G, K. Metgzer, I. Mazzaro, G. Kellermann, E. Ribeiro, J. Varalda, D. H. Mosca, Single-step formation of $Cr_2N$ nanoparticles by pulsed laser irradiation, J. Appl. Phys. 125 (2019) 024301.

[42] H. Era, Y. Ide, A. Nino, K. Kishitake, TEM study on chromium nitride coatings deposited by reactive sputter method, Surf. Coat. Technol. 194 (2005) 265-270.

[43] W. Spengler, R. Kaiser, and H. Bilz, Resonant Raman scattering in a superconducting transition metal compound TiN, Solid State Commun. 17 (1975) 19-22.

[44] M. Stoehr, C. S Shin, I. Petrov, J. E. Greene, Raman scattering from $TiN_x$ ($0.67 \leq x \leq 1.00$) single crystals grown on MgO (001), J. Appl. Phys. 110 (2011) 083503.

[45] K. C. Maurya, B. Biswas, M. Garbretch, Wave-Vector-Dependent Raman Scattering from Coupled Plasmon–Longitudinal Optical Phonon Modes and Fano Resonance in n-type Scandium Nitride, B. Saha, Phys. Status Solidi Rapid Res. Lett. 13 (2019) 1900196.

**Supplementary Information:**

**1. Symmetric ω-2θ X-Ray diffraction (XRD):**

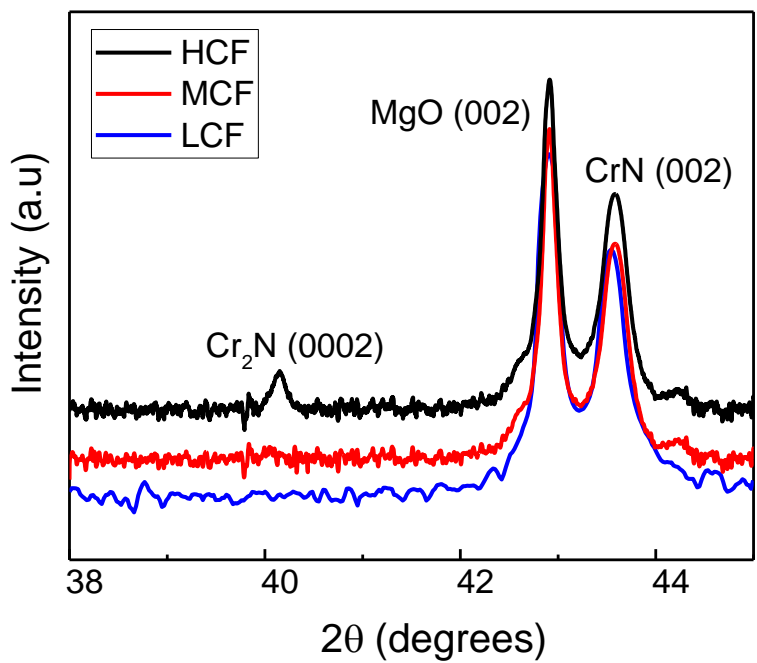

Fig. S1: Symmetric $2\theta - \omega$ HRXRD patterns of the CrN thin films deposited at different Cr flux.

Fig. S1 shows all the films grow with 002 orientation on (002) MgO substates predominantly. For HCF film, in addition to the 002 peak of CrN, 0002 peak of metallic hexagonal $Cr_2N$ is also present, which is absent in both the MCF and LCF films. While the absence of 0002 peak in the LCF film is due to the phase pure single-crystalline nature, the absence of 0002 peak in the MCF film is due to poor scattering intensity and absence of $Cr_2N$ grains along the *c*-axis direction.

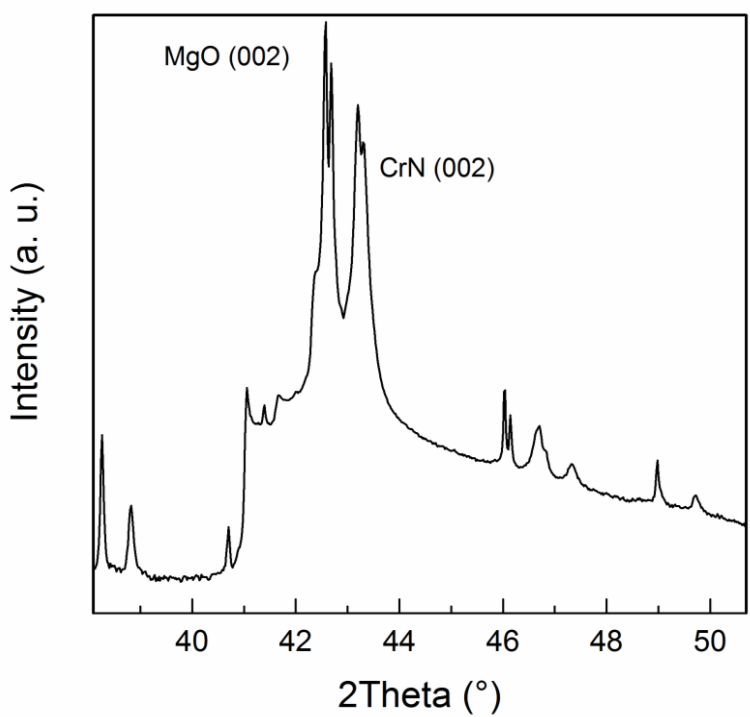

Fig. S2: Powder X-ray diffractogram of the HCF thin film, performed after annealing, is presented. The (002) peak of cubic CrN is present, but the $Cr_2N$ (0002) peak at 40.3°, which was present in the XRD diffractogram taken before annealing (Fig. S1), is absent here. Owing to the nitridation at elevated temperature, $Cr_2N$ grains transform into CrN, and thus $Cr_2N$ (0002) peak disappears.

## 2. Scanning Electron Microscopy (SEM) Imaging:

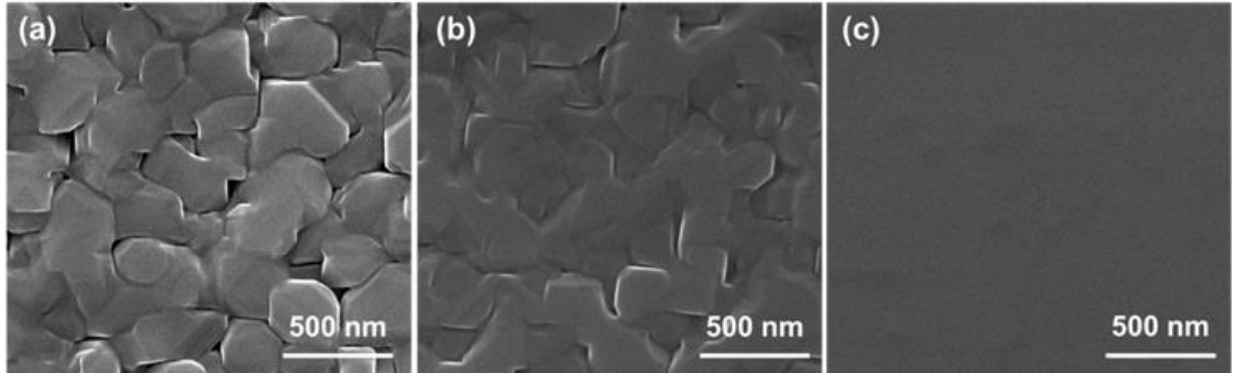

Fig. S3: Plan-view SEM images of (a) HCF, (b) MCF and (c) LCF thin films.

For the HCF film, individual grains of $Cr_2N$ and CrN of sizes 500 - 600 nm can be seen in the plan-view SEM image (Fig. S2 (a)). The grains are separated by voids. For the MCF film, the voids started to disappear and the film looks more compact and smoother compared to HCF film (Fig. S2 (b)). The individual grains of $Cr_2N$ and CrN can be identified in this film also. For the LCF film, which only contains CrN grains, the surface appears compact and smoother without the presence of voids (Fig. S2 (c)).

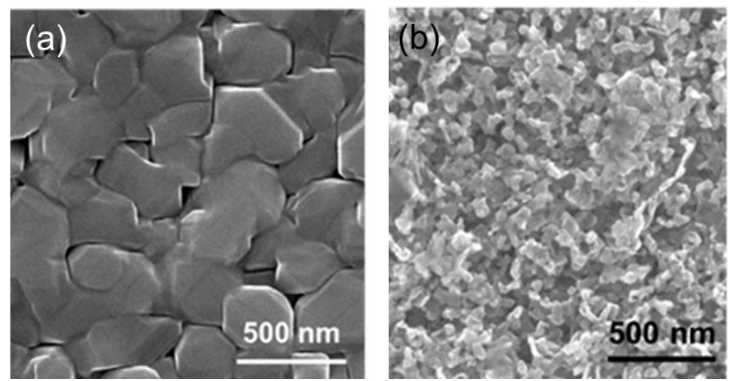

Fig. S4: Plan-view SEM images of HCF film before (a) and after (b) annealing. The film surface looks very rough and porous after annealing. The prolonged annealing at 800°C caused the surface to look significantly different.

## 4. Energy Dispersive X-ray Spectroscopy (EDS):

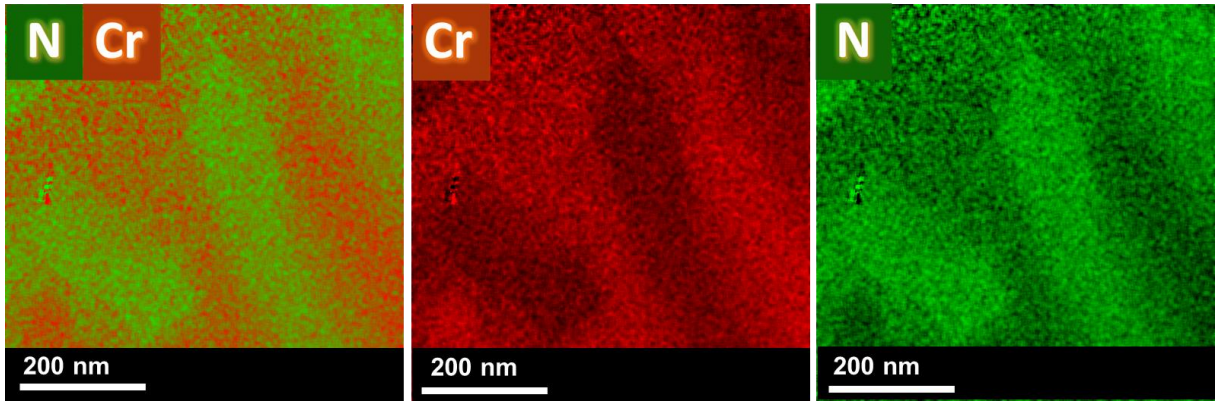

Fig. S5: Elemental mapping of the inhomogeneous regions of the HCF film corresponding to Fig. 2 in the main manuscript at higher magnification showing two distinct phases with different Cr and N concentrations. Due to higher Cr concentration, Cr signal is more intense in the $Cr_2N$ region compared to the CrN region. Similarly, the N signal is more intense in the CrN region compared to the $Cr_2N$ region.

**5. Intensity Variation of $B_{1g}/A_g(1)$ Raman Peak with Temperature:**

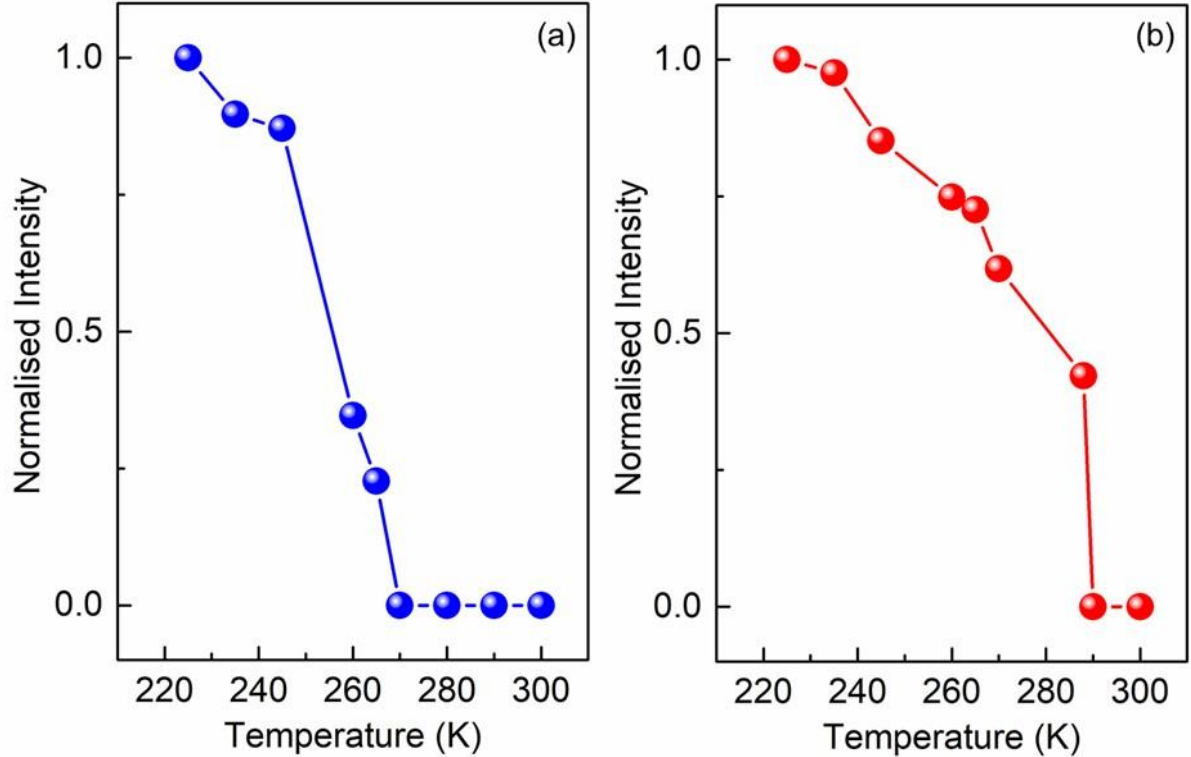

Fig. S6: Intensity variation of $B_{1g}/A_g(1)$ Raman peak with temperature of the HCF film (a) and the LCF film showing the onset of structural transition. Normalized peak intensity has been calculated with respect to maximum intensity in each film. For the HCF film, the structural transition starts around 265 K, and for the LCF film, the structural transition starts around 288 K.

## 6. Raman Peak Intensity Difference between HCF and LCF film

The origin of the higher intensity Raman signals in the LCF film as compared to the HCF film is related to the crystalline quality of the films. The LCF film is a single-phase epitaxial highly-crystalline CrN film having larger grain size, while the HCF film is a mixed-phase polycrystalline $Cr_2N$-CrN lateral heterostructure with smaller sized grains. Hence, from a crystal quality point of view, the HCF film is expected to result in a much smaller Raman cross-section. Peak intensity variation in Raman scattering arises due to the differences in the polarizability of different Raman modes, and polarizability depends on the grain orientation also. The modes at ~ 460 cm-1 and ~ 500 cm-1 are Ag modes, and Ag modes are in general highly dependent on crystal orientation. In the LCF film, the peak at ~ 500 cm-1 is more intense than at ~ 460 cm-1 which clearly highlights the dependence of polarizability on grain orientation.